\documentclass[aps,prd,nofootinbib,superscriptaddress,twocolumn,eqsecnum,floatfix]{revtex4-2}
\usepackage{amsmath,amssymb,amsfonts}
\usepackage{graphicx}
\usepackage{bm}
\usepackage{xcolor}
\usepackage{orcidlink}
\usepackage{hyperref}
\hypersetup{
  colorlinks=true,
  linkcolor=blue,
  citecolor=cyan,
  urlcolor=cyan
}

\newcommand{\dd}{\mathrm{d}}
\newcommand{\rs}{r_s}        
\newcommand{\vs}{v_s}        
\newcommand{\lo}{l_0}         
\newcommand{\el}{\ell}         

\begin{document}

\title{Quantum-gravity-inspired Alcubierre warp-drive geometries}

\author{Kimet Jusufi\,\orcidlink{0000-0003-0527-4177}}
\email{kimet.jusufi@unite.edu.mk}
\affiliation{Physics Department, State University of Tetovo, Ilinden Street nn, 1200, Tetovo, North Macedonia}

\author{Francisco S.N. Lobo\,\orcidlink{0000-0002-9388-8373}}%
\email{fslobo@fc.ul.pt}
\affiliation{Departamento de F\'{i}sica, Faculdade de Ci\^{e}ncias da Universidade de Lisboa, Campo Grande, Edif\'{\i}cio C8, P-1749-016 Lisbon, Portugal}
\affiliation{Instituto de Astrof\'{\i}sica e Ci\^{e}ncias do Espa\c{c}o, Faculdade de Ci\^encias da Universidade de Lisboa, Campo Grande, Edif\'{\i}cio C8, P-1749-016 Lisbon, Portugal}%

\begin{abstract}
String T-duality, through its correspondence with path-integral duality, endows the low-energy propagator with a zero-point length $\lo=2\pi\sqrt{\alpha'}$ that softens the short-distance behavior of gravitational fields. Motivated by the regular black-hole construction obtained from the corresponding smeared source, we formulate a T-duality-inspired Alcubierre geometry by identifying the warp profile with the complementary cumulative mass fraction of that source. For the effective one-scale choice $f(\rs)=1-\rs^3/(\rs^2+\el^2)^{3/2}$, with $\el^2=R^2+\lo^2$, the metric is $C^2$ at the bubble center and smooth elsewhere, while the energy density, its volume integral, and the York expansion are available in closed form. In particular, $E=-(15\pi/1024)\vs^2\el$ in geometric units and $|\rho_E|$ is bounded by a constant times $\vs^2/\el^2$. Within this one-scale family, the model therefore removes the divergence associated with the profile-contraction limit $R\to0$ at fixed velocity, because $\el$ cannot fall below $\lo$. This should not be confused with a regularization of the conventional Alcubierre thin-wall limit at fixed macroscopic bubble radius, in which an independent wall thickness is taken to zero. We emphasize, however, that the construction is an effective ansatz rather than a derivation from string-corrected field equations: $R$ is a macroscopic profile scale, the choice $\el^2=R^2+\lo^2$ is an interpolation, and the classical $\lo\to0$ limit remains a smooth thick-walled profile rather than the distributional Alcubierre top hat. For $\vs>1$, the condition $\vs[1-f]=1$ defines an axial null-characteristic radius and a spherical stationary-limit (Killing-norm-zero) surface of the comoving Killing field, but not a spherical $3+1$ null horizon; accordingly no area entropy is assigned. Exotic matter remains necessary, and neither semiclassical stability nor a modified quantum energy inequality is claimed without an explicit renormalized stress-tensor calculation.
\end{abstract}
\maketitle

\section{Introduction}
\label{sec:intro}
Two of the more dramatic constructions associated with classical general relativity---curvature singularities in black-hole solutions and superluminal ``warp-drive'' geometries---arise in very different physical settings, but both can involve limiting configurations in which the relevant source or profile becomes distributionally sharp. For the Alcubierre spacetime \cite{Alcubierre1994}, the thin-wall limit concentrates the negative Eulerian energy density into an increasingly narrow region, while the quantum-inequality analysis of Pfenning and Ford \cite{PfenningFord1997} drives the admissible wall thickness toward microscopic scales where the semiclassical approximation itself becomes delicate.

A complementary line of reasoning starts from the possibility that quantum gravity supplies an invariant short-distance scale. Padmanabhan's path-integral duality \cite{Padmanabhan1997,Padmanabhan1998} implements an invariance under $l\to\lo^2/l$, producing an ultraviolet-softened propagator. Smailagic, Spallucci, and Padmanabhan \cite{SSP2003} showed that the same structure emerges from the T-duality of closed strings winding a compact dimension, with $\lo=2\pi\sqrt{\alpha'}$. Nicolini, Spallucci, and Wondrak \cite{NSW2019} used the corresponding static potential to construct a regular, neutral black-hole geometry formally identical to the Bardeen solution \cite{Bardeen1968}, with the zero-point length playing the role of the regulator. Related T-duality-inspired geometries have subsequently been explored in charged, rotating, lower-dimensional, cosmological, geodesic completeness, and wormhole settings; see, e.g., Refs.~\cite{Gaete2022,Jusufi:2022rbt,Jusufi:2024dtr,Lobo:2025nng} and references therein.

The purpose of the present work is more modest than deriving a warp drive from string theory. We ask instead whether the same zero-point-length smearing that regularizes a point source can be used as a controlled phenomenological guide for constructing a regular Alcubierre-type profile. The central prescription is to identify the warp shape function with the complementary cumulative mass fraction of the T-duality-smeared source and to introduce an effective profile scale $\el$ satisfying $\el^2=R^2+\lo^2$. This yields a particularly simple one-scale family whose local energy density, volume-integrated negative energy, expansion, and axial causal data can be obtained analytically.

Several qualifications are essential. First, the prescription is an effective ansatz: the smeared warp source is not derived from string-corrected gravitational field equations, and the interpolation $\el^2=R^2+\lo^2$ is chosen for its minimal-length and decoupling properties rather than uniquely dictated by T-duality. Second, the limit $\lo\to0$ of the proposed profile is still smooth and thick-walled; consequently the absence of an arbitrarily thin wall is a property of this one-scale construction and should not be attributed to T-duality alone. Third, exotic matter is not eliminated. Superluminal travel is subject to general negative-energy and energy-condition obstructions \cite{Olum1998,LoboVisser2004}, and more recent analyses stress that positivity of the energy density for a preferred observer congruence is not sufficient to establish the weak or null energy condition \cite{Santiago2022}. A recent systematic classification and critical reassessment of contemporary warp-drive spacetimes, including general no-go statements and cautions concerning physicality claims, is given in Ref.~\cite{Barzegar2026}. Fourth, semiclassical warp-drive backgrounds possess well-known horizon-related pathologies \cite{Hiscock1997,Finazzi2009}; the zero-point-length propagator may alter their ultraviolet structure, but this has to be demonstrated by computing the renormalized stress tensor. Finally, quantum energy inequalities are highly sensitive to the short-distance field theory \cite{FordRoman1996,FewsterEveson1998,KontouSanders2020}; we therefore treat their T-duality deformation as an open problem rather than postulating a specific modified bound.

Within these limits, the main result is an analytically tractable example showing how a minimal profile scale can render the one-scale profile-contraction limit of an Alcubierre-type ansatz finite while leaving the requirement of exotic stress energy untouched. This result concerns the present one-scale family and does not remove the standard fixed-radius Alcubierre thin-wall divergence, where the bubble radius and wall thickness are independent scales. The distinction between regularization, dynamical consistency, and physical realizability will be maintained throughout.

This paper is organized as follows. In Sec.~\ref{sec:review}, we review the zero-point length arising from string T-duality. In Sec.~\ref{sec:construction}, we construct the corresponding regularized Alcubierre geometry and analyze its shape function. Section~\ref{sec:stress} examines the stress-energy distribution, energy bounds, and total negative energy. In Sec.~\ref{sec:kinematics}, we study the expansion and causal structure of the superluminal regime. Section~\ref{sec:quantum} discusses semiclassical backreaction and quantum-energy-inequality constraints as complementary consistency tests of the effective geometry. Finally, in Sec.~\ref{sec:discussion}, we discuss the physical implications, limitations, and possible extensions of the construction and summarize our main conclusions.

Throughout we use geometric units $G=c=1$ unless stated otherwise, and signature $(-,+,+,+)$.

\section{Zero-point length from T-duality: a brief review}
\label{sec:review}

We consider a closed bosonic string propagating in a $(4+1)$-dimensional spacetime with the fifth dimension compactified on a circle of radius $R_5$. The mass spectrum is
\begin{equation}
m^2 = \frac{n^2}{R_5^2} + \frac{w^2 R_5^2}{\alpha'^2} + \frac{2}{\alpha'}\left(N + \tilde N - 2\right),
\end{equation}
Physical closed-string states are additionally subject to the level-matching condition
\begin{equation}
N-\tilde N=nw,
\label{eq:levelmatching}
\end{equation}
which is preserved together with the mass spectrum under the T-duality transformation $R_5\to\alpha'/R_5$, $n\leftrightarrow w$. The latter exchanges Kaluza--Klein momentum modes and winding modes. The self-dual radius $R_5 = \sqrt{\alpha'}$ motivates a characteristic short-distance scale: compactifications with radii $R_5$ and $\alpha'/R_5$ are physically dual, although interpreting this statement as a universal minimum spacetime distance is an additional phenomenological step rather than a general theorem of T-duality. Smailagic, Spallucci, and Padmanabhan \cite{SSP2003} showed by explicit path-integral computation that the center-of-mass propagator of such a string, projected to four dimensions, takes the form
\begin{equation}
G(k) = -\,\frac{\lo\; K_1\!\left(\lo \sqrt{k^2+m^2}\right)}{\sqrt{k^2+m^2}},
\label{eq:propagator}
\end{equation}
with $\lo = 2\pi\sqrt{\alpha'}$, where $K_1$ is a modified Bessel function of the second kind. For $\lo\sqrt{k^2+m^2} \ll 1$ one recovers the standard propagator $\sim (k^2+m^2)^{-1}$, while for large momenta the propagator is exponentially damped: within this construction, the zero-point length acts as an ultraviolet-softening scale or regulator rather than a sharp momentum cutoff. Equation \eqref{eq:propagator} coincides with the propagator postulated in path-integral duality \cite{Padmanabhan1997,Padmanabhan1998}, providing a string-theoretic realization of the same low-energy propagator structure in this setting; see also the published follow-up analysis of Fontanini, Spallucci, and Padmanabhan~\cite{Fontanini2006}.

The static potential between two masses exchanged by virtual gravitons with propagator \eqref{eq:propagator} is \cite{NSW2019}
\begin{equation}
V(r) = -\,\frac{M}{\sqrt{r^2 + \lo^2}},
\label{eq:potential}
\end{equation}
which is finite at $r=0$ and reduces to the Newtonian potential $-M/r$ for $r \gg \lo$. This potential can be interpreted as arising from a smeared source density $\rho_{\lo}(r)$ that is no longer a Dirac delta but a smooth function of width $\lo$. Acting with the Laplacian, $\nabla^2 V = 4\pi \rho_{\lo}$, gives the effective smeared density,
\begin{equation}
\rho_{\lo}(r) = \frac{3 M \lo^2}{4\pi (r^2+\lo^2)^{5/2}},
\label{eq:smeared}
\end{equation}
whose cumulative mass function (the mass enclosed within a sphere of radius $r$) is
\begin{equation}
m(r) \;=\; 4\pi \int_0^r \rho_{\lo}(r')\, r'^2\, \dd r'
\;=\; \frac{M\, r^3}{(r^2+\lo^2)^{3/2}}.
\label{eq:massfunction}
\end{equation}
This function interpolates smoothly from $m(r) \approx M r^3/\lo^3$ at $r \ll \lo$ (a de Sitter-like core) to $m(r) \approx M(1 - 3\lo^2/2r^2)$ at infinity, with the mass completely smeared over the zero-point length scale.

Coupling this source to Einstein's equations in the static, spherically symmetric sector produces the regular metric of Ref.~\cite{NSW2019},
\begin{equation}
-g_{tt} = 1 - \frac{2 M r^2}{(r^2+\lo^2)^{3/2}},
\label{eq:bardeen}
\end{equation}
formally the Bardeen geometry with the magnetic charge replaced by $\lo$. Its curvature invariants are bounded by inverse powers of $\lo$, it possesses a de Sitter core, and its Hawking temperature reaches a maximum before a cold remnant phase---features shared by essentially all minimal-length--inspired black holes, suggesting a universality of quantum corrections \cite{NSW2019}.

Two properties of Eqs.~\eqref{eq:smeared}--\eqref{eq:massfunction} matter for what follows. First, the fraction of the source contained within radius $r$, namely $m(r)/M = r^3/(r^2+\lo^2)^{3/2}$, interpolates smoothly and monotonically from $0$ to $1$. Second, corrections to any classical quantity evaluated at $r \gg \lo$ are of order $\lo^2/r^2$: the zero-point length decouples from macroscopic physics except through its regularizing role. This is a generic feature of zero-point-length modifications, which we will exploit.

For a recent review of zero-point-length phenomenology and its relation to path-integral duality and string T-duality, see Ref.~\cite{Nicolini2022}.

\section{The T-duality--improved warp bubble}
\label{sec:construction}

\subsection{Where the distributional limit hides in a warp drive}

The Alcubierre metric \cite{Alcubierre1994} in $3+1$ ADM form reads
\begin{equation}
\dd s^2=-\dd t^2+\left(\dd x-\vs(t)f(\rs)\dd t\right)^2+\dd y^2+\dd z^2,
\label{eq:alcubierre}
\end{equation}
with $x_s(t)$ the bubble trajectory, $\vs=\dd x_s/\dd t$, and $\rs=[(x-x_s)^2+y^2+z^2]^{1/2}$. The lapse is unity, and the geometry is encoded in the shape function $f$, with $f(0)=1$ and $f\to0$ at spatial infinity. Alcubierre's original profile is
\begin{equation}
f_{\rm A}(\rs)=\frac{\tanh[\sigma(\rs+R)]-\tanh[\sigma(\rs-R)]}{2\tanh(\sigma R)},
\label{eq:tanh}
\end{equation}
which approaches $\Theta(R-\rs)$ as the inverse wall thickness $\sigma\to\infty$. Since the Eulerian energy density is quadratic in $f'$ [Eq.~\eqref{eq:rhoE}], a wall of thickness $\Delta$ has schematically $f'\sim\Delta^{-1}$, $\rho_E\sim\Delta^{-2}$, and $|E|\sim\Delta^{-1}$. The limiting energy density is therefore more singular than an ordinary delta distribution and should not itself be identified with a Dirac delta.

The relevance of this regime is physical rather than merely formal. Pfenning and Ford \cite{PfenningFord1997} showed that, within the assumptions of their quantum-inequality analysis, a macroscopic Alcubierre bubble moving with $\vs\sim1$ requires an extremely thin wall and correspondingly enormous negative energy. Thus the regime in which semiclassical constraints become strongest is also the regime in which a putative zero-point length cannot automatically be neglected.

It is important to distinguish this standard thin-wall limit from the limit studied below. In Eq.~\eqref{eq:tanh}, the macroscopic bubble radius and the wall thickness are independent: one may hold the former fixed while sending the latter to zero. In the one-scale profile introduced below, by contrast, the effective scale $\el$ controls both the location and the width of the transition. The limit $R\to0$ therefore contracts the entire one-scale configuration toward its minimum scale $\lo$; it is not the fixed-radius thin-wall limit of the usual Alcubierre family.

\subsection{The smeared shape function}

In the static black-hole problem, the T-duality potential can be represented by the smooth density \eqref{eq:smeared} and cumulative mass \eqref{eq:massfunction}. We use that cumulative profile as a phenomenological template for the warp function and set
\begin{equation}
f(\rs)=1-\frac{m(\rs)}{M}
=1-\frac{\rs^3}{(\rs^2+\el^2)^{3/2}},
\label{eq:shape}
\end{equation}
where
\begin{equation}
\el^2\equiv R^2+\lo^2.
\label{eq:ell}
\end{equation}
Here $R$ should be understood as a macroscopic profile scale rather than, strictly speaking, the radius of a top-hat bubble. The quadrature prescription in Eq.~\eqref{eq:ell} is an effective interpolation: it guarantees $\el\ge\lo$, gives an even expansion in $\lo/R$, and reduces to $\el\simeq R[1+\lo^2/(2R^2)+\cdots]$ for $R\gg\lo$. It is not uniquely derived from the T-duality propagator.

This distinction also clarifies the classical limit. Sending $\lo\to0$ gives
\begin{equation}
f_0(\rs)=1-\frac{\rs^3}{(\rs^2+R^2)^{3/2}},
\label{eq:classicalprofile}
\end{equation}
which remains a smooth, intrinsically thick profile and does not reduce to Eq.~\eqref{eq:tanh} in its sharp-wall limit. Accordingly, the locking of wall location and wall thickness is a property of the present one-scale ansatz. A more microscopic construction would instead smear a finite-radius shell or the characteristic function $\Theta(R-\rs)$ with the zero-point-length kernel; such a construction could preserve an independent macroscopic radius while generating an irreducible quantum wall thickness. We leave that problem for future work.

\begin{figure}[t]
    \centering
    \includegraphics[width=1.0\linewidth]{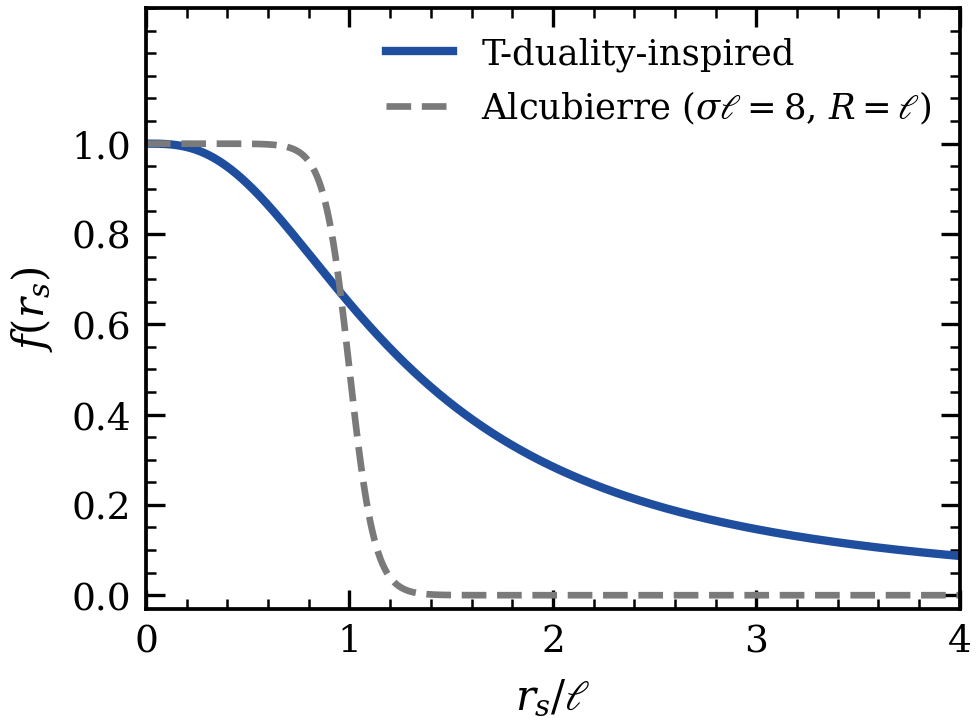}
    \caption{Shape functions $f(\rs)$ as functions of $\rs/\el$. The solid blue curve is the T-duality-inspired profile of Eq.~\eqref{eq:shape},
    $f(\rs)=1-\rs^3/(\rs^2+\el^2)^{3/2}$,
    and the dashed gray curve is the Alcubierre $\tanh$ profile of Eq.~\eqref{eq:tanh},
    $f_{\rm A}(\rs)=\left\{\tanh[\sigma(\rs+R)]-\tanh[\sigma(\rs-R)]\right\}/\left[2\tanh(\sigma R)\right]$,
    evaluated for $\sigma\el=8$ and $R=\el$. The comparison is illustrative: the present one-scale ansatz remains smooth even when $\lo\to0$, whereas the independent parameter $\sigma$ permits the standard Alcubierre profile to approach a step function.}
    \label{fig:wd1}
\end{figure}

\begin{figure}[t]
    \centering
    \includegraphics[width=1.0\linewidth]{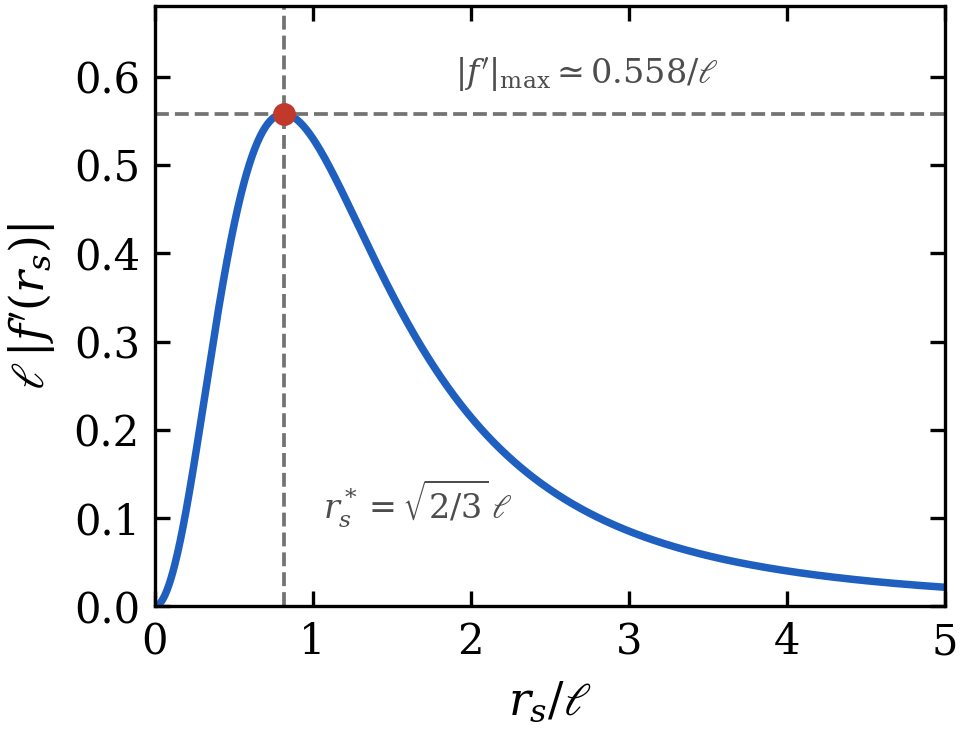}
    \caption{Wall profile $\el|f'(\rs)|$ for the fiducial $n=1$ geometry, showing the peak at $\rs^*=\sqrt{2/3}\,\el$ and $|f'|_{\max}=2(3/5)^{5/2}/\el\simeq0.558/\el$. Since $\el\ge\lo$, the slope is bounded within the adopted family.}
    \label{fig:wd2}
\end{figure}

The profile \eqref{eq:shape} satisfies $f(0)=1$, decreases monotonically, and is sufficiently differentiable to yield finite curvature.\footnote{The profile is analytic for $\rs>0$. At the bubble center it is $C^2$ but not $C^3$, since $\rs^3=(x^2+y^2+z^2)^{3/2}$ has discontinuous third derivatives there. The metric is therefore $C^2$ and the Riemann tensor is continuous at $\rs=0$, although derivatives of curvature need not be.} Its asymptotic form is
\begin{equation}
f(\rs)=\frac{3\el^2}{2\rs^2}-\frac{15\el^4}{8\rs^4}+O(\el^6/\rs^6),
\qquad \rs\gg\el,
\label{eq:falloff}
\end{equation}
which is sufficient for asymptotic flatness and convergence of the energy integral below. Its derivative is
\begin{equation}
f'(\rs)=-\frac{3\el^2\rs^2}{(\rs^2+\el^2)^{5/2}},
\label{eq:fprime}
\end{equation}
with maximum magnitude at
\begin{equation}
\rs^\ast=\sqrt{\frac{2}{3}}\,\el,
\qquad
|f'(\rs^\ast)|=2\left(\frac35\right)^{5/2}\frac1\el
\simeq\frac{0.558}{\el}.
\label{eq:fprimemax}
\end{equation}
Thus the local slope is bounded by $0.558/\lo$ within the family. Figure~\ref{fig:wd1} compares the resulting profile with the conventional Alcubierre function, while Fig.~\ref{fig:wd2} displays the dimensionless wall slope.

An alternative profile obtained directly from the regular potential is $f_V(\rs)=\el/(\rs^2+\el^2)^{1/2}$. It shares the same qualitative regularity but decays only as $\el/\rs$; we do not pursue it here.

\subsection{Generalized shape-function family}
\label{sec:family}

A useful extension is
\begin{equation}
f_n(\rs)=1-\frac{\rs^{2n+1}}{(\rs^2+\el^2)^{(2n+1)/2}},
\qquad n=1,2,3,\ldots,
\label{eq:family}
\end{equation}
with
\begin{equation}
f_n'(\rs)=-\frac{(2n+1)\el^2\rs^{2n}}{(\rs^2+\el^2)^{(2n+3)/2}}.
\label{eq:fnprime}
\end{equation}
The case $n=1$ is Eq.~\eqref{eq:shape}. Substitution into Eq.~\eqref{eq:Edef} gives
\begin{equation}
E_n=-\frac{\vs^2\el}{24}(2n+1)^2
\frac{\Gamma(2n+\tfrac32)\Gamma(\tfrac32)}{\Gamma(2n+3)},
\label{eq:Efamily}
\end{equation}
which reduces to Eq.~\eqref{eq:totalenergy} at $n=1$.

The large-$n$ behavior requires some care. At fixed $\el$,
\begin{equation}
f_n(\rs)=1-\left(1+\frac{\el^2}{\rs^2}\right)^{-(2n+1)/2}
\longrightarrow 1
\end{equation}
for every fixed $\rs>0$, because the transition region moves outward. The maximum of $|f_n'|$ occurs at
\begin{equation}
r_n^\ast=\el\sqrt{\frac{2n}{3}},
\end{equation}
and Stirling's formula gives
\begin{equation}
E_n\sim-\frac{\sqrt{\pi}}{24\sqrt2}\,\vs^2\el\sqrt n,
\qquad n\to\infty.
\label{eq:Efamilyasympt}
\end{equation}
Thus the family does not approach a top hat at fixed $\el$.

As a purely mathematical reparametrization, one may instead set $\el_n=r_\star\sqrt{3/(2n)}$ so that the wall-slope peak $r_n^\ast=r_\star$ remains fixed. In that case the limiting profile is still smooth:
\begin{equation}
\lim_{n\to\infty}f_n(\rs;\el_n)
=1-\exp\!\left(-\frac{3r_\star^2}{2\rs^2}\right),
\label{eq:familylimit}
\end{equation}
and the energy tends to the finite value
\begin{equation}
\lim_{n\to\infty}E_n(\el_n)
=-\frac{\sqrt{3\pi}}{48}\,\vs^2r_\star.
\label{eq:Efamilylimit}
\end{equation}
This $n\to\infty$ construction should not, however, be interpreted as a limit within the physical T-duality-inspired family at fixed nonzero $\lo$. Indeed, the minimum-length condition $\el_n\ge\lo$ implies
\begin{equation}
n\leq \frac{3r_\star^2}{2\lo^2},
\label{eq:nmax}
\end{equation}
so an arbitrarily large-$n$ peak-fixed sequence eventually violates the assumed lower bound on the profile scale. Equations~\eqref{eq:familylimit}--\eqref{eq:Efamilylimit} are therefore mathematical limiting statements for the generalized shape functions (or, equivalently, require a simultaneous $\lo\to0$ limit), not fixed-$\lo$ T-duality limits. Within their proper domain they still demonstrate that this generalized family does not conceal a distributional top-hat limit under the rescalings considered here.

\section{Stress-energy, energy bounds, and total energy}
\label{sec:stress}

\subsection{Energy density}

For any metric of the form \eqref{eq:alcubierre}, the energy density measured by the Eulerian (normal) observers $n^\mu = (1, \vs f, 0, 0)$ is \cite{Alcubierre1994}
\begin{equation}
\rho_E \;\equiv\; T_{\mu\nu} n^\mu n^\nu
\;=\; -\,\frac{\vs^2}{32\pi}\,\frac{y^2+z^2}{\rs^2}\left(\frac{\dd f}{\dd \rs}\right)^{\!2}.
\label{eq:rhoE}
\end{equation}
This expression is derived from the Einstein equations and reflects the fact that the warp drive requires negative energy density. Inserting Eq.~\eqref{eq:fprime} and writing $y^2+z^2 = \rs^2\sin^2\theta$ (with $\theta$ the polar angle from the direction of motion),
\begin{equation}
\rho_E(\rs,\theta) \;=\; -\,\frac{9\,\vs^2}{32\pi}\;
\frac{\el^4\, \rs^4 \sin^2\theta}{\left(\rs^2+\el^2\right)^{5}} .
\label{eq:density}
\end{equation}
For the present profile this quantity is manifestly non-positive, with strict negativity off the symmetry axis wherever $f'\neq0$. Hence the weak energy condition already fails for the Eulerian observers; this explicit result is consistent with the broader superluminal and Alcubierre/Nat\'ario energy-condition obstructions of Refs.~\cite{Olum1998,LoboVisser2004}. It is finite everywhere, exactly zero at the bubble center and along the axis of motion, maximal in magnitude on the equator of the wall at the same radius $\rs^\ast = \sqrt{2/3}\,\el$ as in Eq.~\eqref{eq:fprimemax}, and decays as $\rs^{-6}$. Its extremal value is
\begin{equation}
\big|\rho_E\big|_{\max} = \frac{\vs^2}{8\pi}\left(\frac{3}{5}\right)^{5}\frac{1}{\el^2}
\;\le\; \frac{1}{8\pi}\left(\frac{3}{5}\right)^{5}\frac{\vs^2}{\lo^2},
\label{eq:rhobound}
\end{equation}
where the last inequality follows from $\el\ge\lo$. Thus, within the adopted family and at fixed $\vs$, shrinking the macroscopic profile scale $R$ cannot make the Eulerian density diverge: the limit saturates at $\el=\lo$. The same statement applies to the curvature associated with the profile-width limit. Since the metric is $C^2$ and $f'$, $f''$ are finite, the Riemann tensor is finite everywhere. For constant $\vs$, its nonzero components scale dimensionally as combinations of $\vs f''\sim\vs/\el^2$ and $\vs^2(f')^2\sim\vs^2/\el^2$, so quadratic curvature invariants are bounded at fixed $\vs$ by profile-dependent combinations of order $\vs^2/\el^4$ and $\vs^4/\el^4$. For a time-dependent trajectory, additional terms proportional to time derivatives of $\vs$ occur; these remain finite provided the trajectory is sufficiently smooth and its acceleration is bounded. This is a regularization of the one-scale profile-width limit, not a proof that arbitrary high-velocity or arbitrarily accelerated limits are bounded.

\begin{figure}[t]
    \centering
    \includegraphics[width=1.0\linewidth]{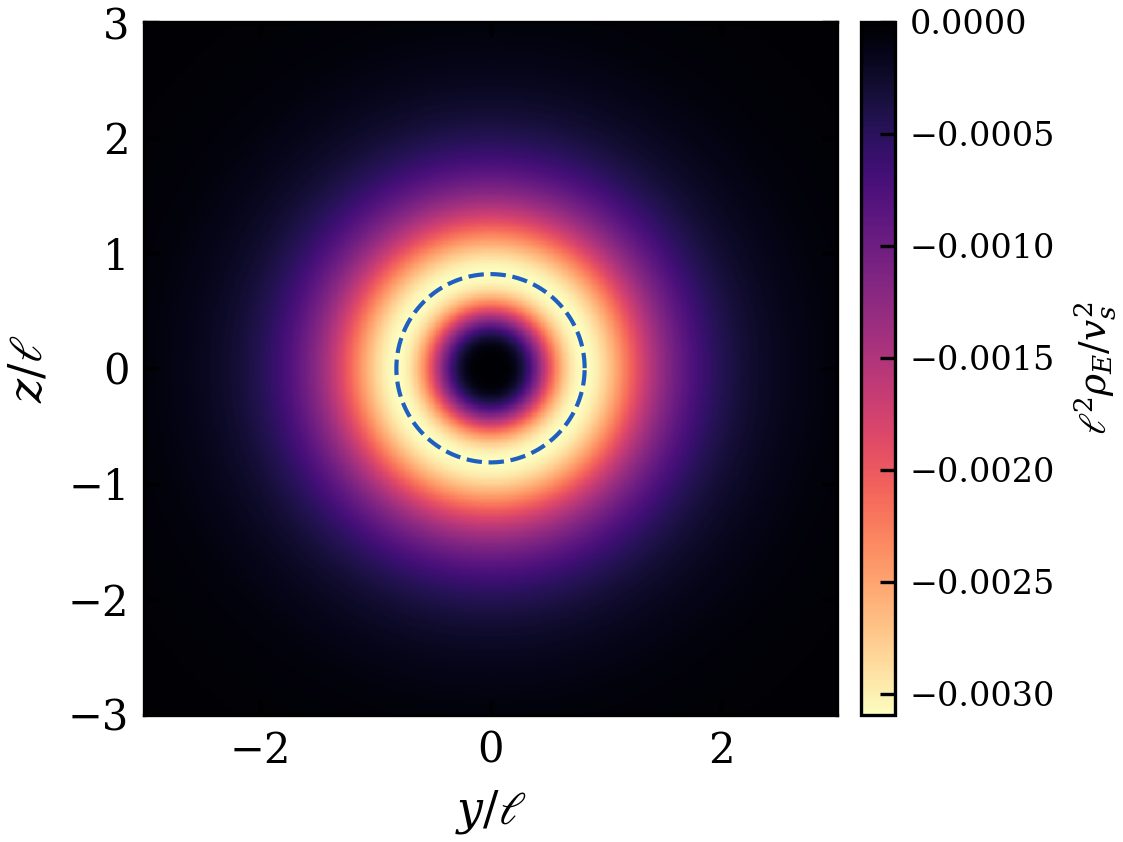}
    \caption{Eulerian energy density $\el^2\rho_E/\vs^2$ in the transverse ($y$-$z$) plane through the bubble center, $\Delta x=0$, where $\sin\theta=1$ and $\rho_E=-(9\vs^2/32\pi)\,\el^4\rs^4/(\rs^2+\el^2)^5$. The negative-energy density has a toroidal concentration; the dashed circle marks the maximum at $\rs^*=\sqrt{2/3}\,\el$, and the magnitude is bounded by Eq.~\eqref{eq:rhobound}.}
    \label{fig:wd3}
\end{figure}

Restoring units, Eq.~\eqref{eq:rhobound} reads $|\rho_E|_{\max} = (3/5)^5 (\vs/c)^2 c^4 / (8\pi G \el^2)$; for a bubble with $\el = 100\,$m at $\vs = c$ this is about $3.7\times 10^{37}\,\mathrm{J\,m^{-3}}$ of negative energy density---enormous, but finite and, notably, spread over a wall of macroscopic thickness rather than concentrated on a Planck-width shell. The toroidal concentration of the transverse negative-energy density is displayed in Fig.~\ref{fig:wd3}.

\subsection{Total energy}

The total Eulerian energy on a flat slice ($\sqrt{h}=1$ for the metric \eqref{eq:alcubierre}) is
\begin{align}
E &= \int \rho_E \,\dd^3x
= -\frac{\vs^2}{32\pi}\int_0^\infty \dd\rs\, \rs^2 \big(f'\big)^2 \int \dd\Omega\, \sin^2\theta
\nonumber\\
&= -\frac{\vs^2}{12}\int_0^\infty \dd\rs\, \rs^2 \big(f'\big)^2 ,
\label{eq:Edef}
\end{align}
using $\int \dd\Omega \sin^2\theta = 8\pi/3$. With Eq.~\eqref{eq:fprime},
\begin{align}
\int_0^\infty \rs^2 (f')^2\, \dd\rs
&= 9\el^4 \int_0^\infty \frac{\rs^6\, \dd\rs}{(\rs^2+\el^2)^{5}}.
\label{eq:Eintegral}
\end{align}
With the substitution $\rs = \el\, t$, and with $B$ the Euler beta function,
\begin{align}
\int_0^\infty \frac{\rs^6\, \dd\rs}{(\rs^2+\el^2)^5}
&= \frac{1}{\el^3}\int_0^\infty \frac{t^6\, \dd t}{(1+t^2)^5}
= \frac{1}{2\el^3}\, B\!\left(\frac{7}{2},\frac{3}{2}\right)
\nonumber\\
&= \frac{1}{2\el^3}\cdot\frac{\Gamma(7/2)\,\Gamma(3/2)}{\Gamma(5)}
= \frac{5\pi}{256\,\el^3},
\end{align}
using $\Gamma(7/2) = 15\sqrt{\pi}/8$, $\Gamma(3/2)=\sqrt{\pi}/2$, $\Gamma(5)=24$. Hence
$\int_0^\infty \rs^2 (f')^2 \dd\rs = 9\el^4 \cdot 5\pi/(256\,\el^3) = 45\pi\el/256$, which provides the exact, closed-form result
\begin{equation}
E \;=\; -\,\frac{15\pi}{1024}\; \vs^2\, \el
\;\approx\; -\,0.046\, \vs^2\, \el .
\;
\label{eq:totalenergy}
\end{equation}
Note that Eq.~\eqref{eq:totalenergy} is the volume integral of the Eulerian energy density on a constant-$t$ slice, which is the standard measure of the ``negative energy requirement'' in the warp-drive literature; it is not an asymptotic conserved charge. Indeed, the spatial slices of Eq.~\eqref{eq:alcubierre} are exactly flat, so the ADM energy of the geometry vanishes identically. This distinction between the volume-integrated Eulerian density and the ADM mass in warp-drive spacetimes is discussed in detail in Ref.~\cite{Schuster2023}.

In SI units, $E = -\frac{15\pi}{1024}\,(\vs/c)^2\, (c^4/G)\,\el$. For $\el = 100\,$m and $\vs = c$ this gives $E \approx -5.6\times 10^{44}\,$J, i.e., about $-0.003\,M_\odot c^2$ (roughly three Jupiter masses of negative energy). This remains an enormous negative-energy budget. Its reduction relative to the Pfenning--Ford thin-wall estimate \cite{PfenningFord1997} is primarily geometric: the present ansatz is intrinsically thick-walled. This is consistent with the general observation that integrated warp-drive energy requirements depend sensitively on the chosen profile \cite{VanDenBroeck1999}. It should therefore not be interpreted as a T-duality loophole in the energy conditions or in quantum energy inequalities.

Expanding Eq.~\eqref{eq:totalenergy} in the macroscopic regime $R\gg\lo$,
\begin{equation}
E(R) = -\frac{15\pi}{1024}\,\vs^2 R \left(1 + \frac{\lo^2}{2R^2} + O(\lo^4/R^4)\right),
\label{eq:Eexpansion}
\end{equation}
the first correction generated by the interpolation \eqref{eq:ell} enters at relative order $\lo^2/R^2$. The same scaling occurs in the regular black-hole observables of Ref.~\cite{NSW2019}. In the present model, however, this behavior is partly built into the choice $\el^2=R^2+\lo^2$; it should therefore be regarded as consistency with, rather than an independent proof of, a universal zero-point-length expansion.

\subsection{The minimal-profile-scale limit}

The limiting case $R\to0$ provides a useful regularity check of the effective interpolation \eqref{eq:ell}:
\begin{equation}
\lim_{R\to0}\el=\lo.
\label{eq:ellmin}
\end{equation}
At fixed $\vs$ the local density and the magnitude of the volume-integrated negative energy therefore approach finite values,
\begin{align}
|\rho_E|_{\max}&\longrightarrow
\frac{\vs^2}{8\pi}\left(\frac35\right)^5\frac1{\lo^2},\\
|E|&\longrightarrow\frac{15\pi}{1024}\vs^2\lo.
\end{align}
It is preferable to describe the latter as the minimum of $|E|$ within this one-scale family,
\begin{equation}
|E|_{\min}=\frac{15\pi}{1024}\vs^2\lo,
\label{eq:Emin}
\end{equation}
rather than as a minimum of the signed quantity $E<0$. If $\lo$ happens to be of order the Planck length, the corresponding mass scale is of order $10^{-9}\,$kg for $\vs\sim1$, but the relation between $\lo=2\pi\sqrt{\alpha'}$ and $\ell_p$ is model dependent.

No dynamical ground state, remnant, soliton, or ultraviolet completion follows from Eq.~\eqref{eq:Emin}. Those notions require an action principle, perturbative stability, or a quantum-gravitational completion that is not supplied by the present construction. The correct conclusion is narrower: the effective family has a finite minimal-width endpoint and no divergence associated with $R\to0$ at fixed velocity.

\begin{figure}[t]
    \centering
    \includegraphics[width=1.0\linewidth]{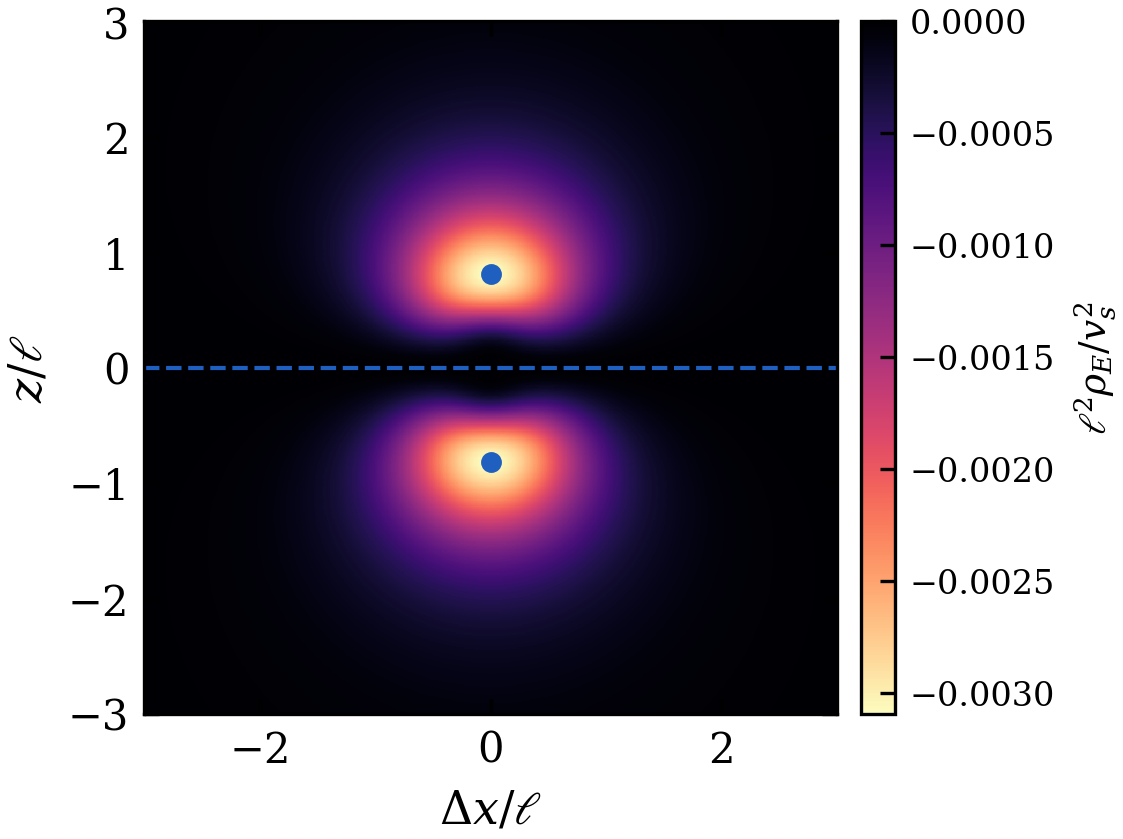}
    \caption{Eulerian energy density $\el^2\rho_E/\vs^2$ in the meridional ($x$-$z$) plane, $y=0$, where $\sin^2\theta=z^2/\rs^2$ and $\rho_E=-(9\vs^2/32\pi)\el^4\rs^2z^2/(\rs^2+\el^2)^5$. This is the cross-section of the toroidally concentrated high-magnitude region in Fig.~\ref{fig:wd3}: the density vanishes on the axis of motion and reaches its extrema at $\Delta x=0$, $z=\pm\sqrt{2/3}\,\el$.}
    \label{fig:wd4}
\end{figure}

Figure~\ref{fig:wd4} shows a meridional section of the Eulerian density and makes explicit that the region of largest negative-energy magnitude is toroidally concentrated rather than spherically distributed; the density itself has noncompact, rapidly decaying support.

\section{Kinematics: expansion and axial causal structure}
\label{sec:kinematics}

\subsection{York expansion}

\begin{figure}[t]
    \centering
    \includegraphics[width=1.0\linewidth]{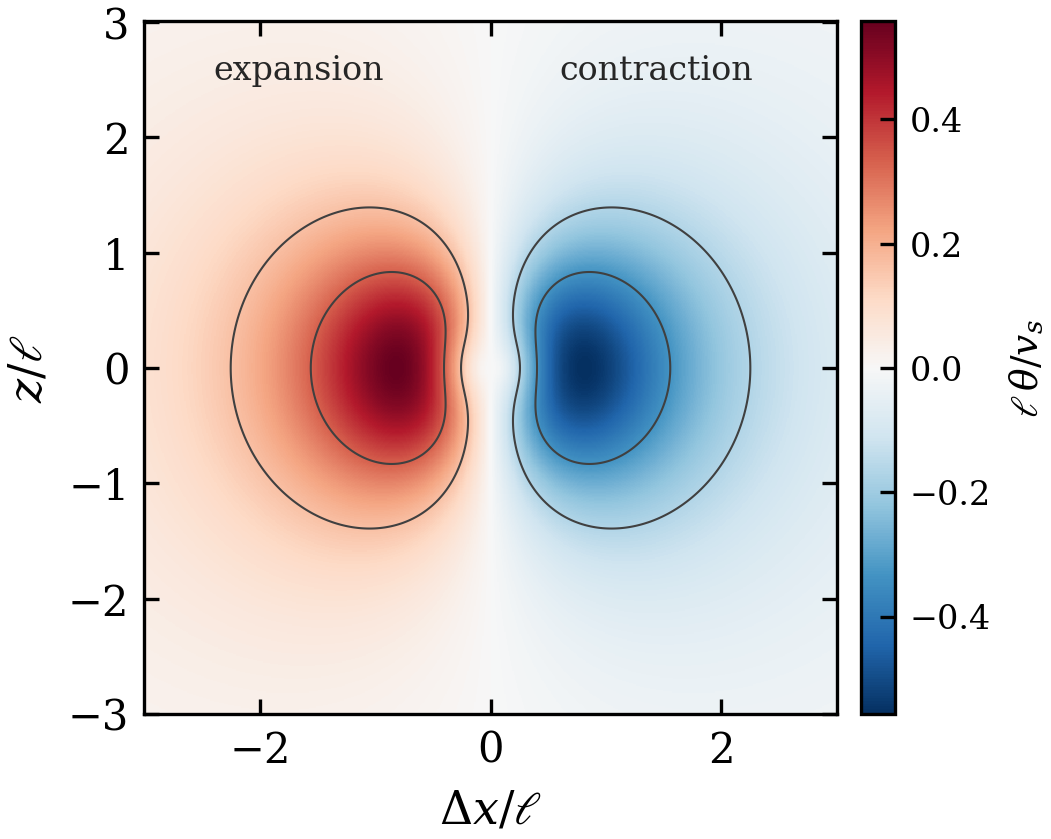}
    \caption{York expansion $\el\theta/\vs$, Eq.~\eqref{eq:expansion}, in the meridional plane. The dipolar contraction/expansion pattern is smooth, and its extrema scale as $O(\vs/\el)$.}
    \label{fig:wd5}
\end{figure}

The expansion of the Eulerian congruence for Eq.~\eqref{eq:alcubierre} is \cite{Alcubierre1994}
\begin{equation}
\theta=\vs\frac{\Delta x}{\rs}f'(\rs),
\qquad \Delta x=x-x_s(t),
\end{equation}
and hence
\begin{equation}
\theta(\rs,\Delta x)=-\frac{3\vs\el^2\rs\Delta x}{(\rs^2+\el^2)^{5/2}}.
\label{eq:expansion}
\end{equation}
The familiar dipolar pattern of contraction ahead of the bubble and expansion behind it is retained, but its magnitude is finite and scales as $\vs/\el$ at fixed velocity. Figure~\ref{fig:wd5} displays this pattern. The statement is one of geometrical regularization; no quantization of the expansion scalar is implied.

\subsection{Stationary-limit surface and axial null characteristics}

For constant $\vs$, introduce the coordinate $X=x-x_s(t)$, so that $\dd x=\dd X+\vs\dd t$. Equation~\eqref{eq:alcubierre} becomes
\begin{equation}
\dd s^2=-\dd t^2+
\left[\dd X+\vs(1-f(\rs))\dd t\right]^2
+\dd y^2+\dd z^2,
\label{eq:comoving}
\end{equation}
where $\rs=(X^2+y^2+z^2)^{1/2}$. Define
\begin{equation}
\beta(\rs)=\vs[1-f(\rs)]
=\vs\frac{\rs^3}{(\rs^2+\el^2)^{3/2}}.
\label{eq:beta}
\end{equation}
For constant $\vs$, the comoving stationary Killing field $\xi=\partial_t$ has norm $\xi^2=g_{tt}=-(1-\beta^2)$. For $\vs>1$ this Killing field is spacelike asymptotically, since $\beta\to\vs$ as $\rs\to\infty$; consequently the zero of its norm should not be interpreted as a Kerr-like ergosurface associated with an asymptotically timelike stationary Killing vector. Therefore the equation
\begin{equation}
\beta(r_{\rm SL})=1
\qquad\Longleftrightarrow\qquad
\vs\frac{r_{\rm SL}^3}{(r_{\rm SL}^2+\el^2)^{3/2}}=1
\label{eq:horizon}
\end{equation}
defines a spherical Killing-norm-zero, or stationary-limit, surface of the comoving Killing field whenever $\vs>1$. Its radius is
\begin{equation}
r_{\rm SL}=\frac{\el}{\sqrt{\vs^{2/3}-1}}.
\label{eq:rh_exact}
\end{equation}
The notation $r_{\rm SL}$ emphasizes that this is a stationary-limit radius and not, by itself, a global horizon radius.

Indeed, the surface $\rs=\mathrm{const}$ is not generally null. From the inverse of Eq.~\eqref{eq:comoving},
\begin{equation}
g^{\mu\nu}(\partial_\mu\rs)(\partial_\nu\rs)
=1-\beta^2\frac{X^2}{\rs^2}
=1-\beta^2\cos^2\vartheta,
\label{eq:normalnorm}
\end{equation}
where $\vartheta$ is the angle from the direction of motion. On $\beta=1$ this reduces to
\begin{equation}
g^{\mu\nu}(\partial_\mu\rs)(\partial_\nu\rs)=\sin^2\vartheta,
\label{eq:stationarynotnull}
\end{equation}
which vanishes only on the symmetry axis. Thus the sphere \eqref{eq:rh_exact} is the stationary-limit (Killing-norm-zero) surface of the comoving Killing field, not a spherical $3+1$ null horizon.

Along the symmetry axis the metric reduces to the $1+1$ Painlev\'e--Gullstrand form
\begin{equation}
\dd s^2_{\rm ax}=-\dd t^2+(\dd X+\beta\dd t)^2,
\end{equation}
whose null characteristics obey
\begin{equation}
\frac{\dd X}{\dd t}=-\beta\pm1.
\label{eq:axialnull}
\end{equation}
The outgoing branch has zero coordinate velocity at $\beta=1$. Equation~\eqref{eq:rh_exact} therefore gives the location at which an axial null characteristic becomes stationary in the comoving coordinates. The continuation of this axial causal boundary to a genuine $3+1$ null hypersurface requires solving the corresponding null-surface equation and is not attempted here.

Because Eq.~\eqref{eq:rh_exact} specifies a radius, the symmetry axis contains two such points, $X_\pm=\pm r_{\rm SL}$. For the branch $\dd X/\dd t=1-\beta$, linearizing about these points gives
\begin{equation}
\frac{\dd\,\delta X}{\dd t}\simeq
\begin{cases}
-\,\beta'(r_{\rm SL})\,\delta X, & X=+{r_{\rm SL}},\\[2pt]
+\,\beta'(r_{\rm SL})\,\delta X, & X=-{r_{\rm SL}},
\end{cases}
\label{eq:axialpeeling}
\end{equation}
so the two axial points have opposite linearized peeling character. This front/rear asymmetry is consistent with the horizon-like causal regions found in the null-geodesic analysis of the Alcubierre geometry by Clark, Hiscock, and Larson~\cite{Clark1999}.

\begin{figure}[t]
    \centering
    \includegraphics[width=1.0\linewidth]{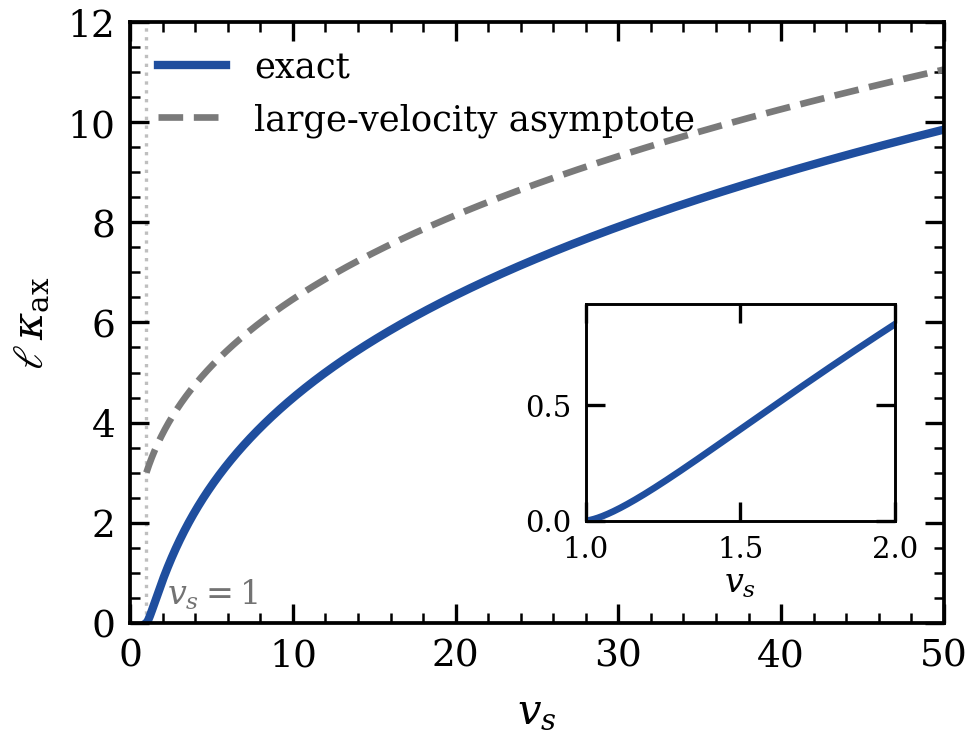}
    \caption{Axial peeling coefficient in units of $1/\el$. The solid blue curve is the exact result of Eq.~\eqref{eq:kappa_exact},
    $\el\kappa_{\rm ax}=3(\vs^{2/3}-1)^{3/2}/\vs^{2/3}$,
    and the dashed gray curve is its large-velocity asymptote, $\el\kappa_{\rm ax}\simeq3\vs^{1/3}$. The inset magnifies the near-threshold region, where $\kappa_{\rm ax}\to0$ as $\vs\to1^+$. This is an axial causal diagnostic, not the surface gravity of a demonstrated spherical $3+1$ horizon.}
    \label{fig:wd6}
\end{figure}

A useful local quantity is the axial peeling coefficient
\begin{equation}
\kappa_{\rm ax}\equiv
\left|\frac{\dd\beta}{\dd\rs}\right|_{r_{\rm SL}}
=\frac{3}{\el}
\frac{(\vs^{2/3}-1)^{3/2}}{\vs^{2/3}}.
\label{eq:kappa_exact}
\end{equation}
It vanishes as $\vs\to1^+$ and behaves as $3\vs^{1/3}/\el$ for $\vs\gg1$. Hence the profile-width regularization bounds $\kappa_{\rm ax}$ at fixed $\vs$ but does not produce a velocity-independent upper bound. Figure~\ref{fig:wd6} illustrates this behavior.

For comparison with the usual $1+1$ near-horizon language one may define the purely formal scale
\begin{equation}
T_{\rm ax}^{\rm formal}\equiv\frac{\kappa_{\rm ax}}{2\pi}.
\label{eq:Thorizon}
\end{equation}
We do not interpret Eq.~\eqref{eq:Thorizon} as the temperature of a $3+1$ warp horizon. A thermal interpretation requires a quantum state, a genuine causal horizon, and a Hadamard/RSET analysis. Existing semiclassical studies show that superluminal warp geometries can develop severe horizon-related stress-energy effects \cite{Hiscock1997,Finazzi2009}.

\subsection{Why no horizon area or entropy is assigned}

Equation~\eqref{eq:stationarynotnull} prevents the stationary-limit sphere from being treated as a null horizon with area $4\pi {\color{blue}r_{\rm SL}}^2$. Consequently an area-law entropy $A/4G$ is not assigned in the present analysis. Establishing a meaningful entropy would first require identifying a genuine global or quasi-local horizon and specifying the gravitational theory and quantum state to which an entropy law applies. The omission is therefore geometrical rather than merely interpretational.

\section{Semiclassical consistency and quantum energy inequalities}
\label{sec:quantum}

The regularity of the classical effective geometry does not by itself establish its semiclassical consistency. Two related questions must be distinguished: whether quantum backreaction remains under control on the background, and whether an admissible quantum state can support the required negative-energy distribution. We address these issues in turn.

\subsection{Semiclassical backreaction}

A full stability analysis requires the renormalized stress-energy tensor in a specified quantum state together with the corresponding semiclassical Einstein equation. Dimensional analysis alone cannot establish stability. If $\el$ is the only local curvature length and $\vs$ is held fixed, one may write schematically
\begin{equation}
\langle T_{\mu\nu}\rangle_{\rm ren}
=\frac{\hbar}{\el^4}\,
{\cal F}_{\mu\nu}\!\left(\vs,\frac{x^\alpha}{\el};\,\text{state}\right),
\label{eq:RSETscale}
\end{equation}
where the dimensionless tensor ${\cal F}_{\mu\nu}$ may become large in special quantum states or near causal accumulation surfaces. Away from such enhancements, the induced metric correction has the parametric magnitude
\begin{equation}
\frac{\delta g}{g}\sim
\frac{\ell_p^2}{\el^2}\,{\cal F},
\label{eq:backreaction}
\end{equation}
so that macroscopic configurations with $\el\gg\ell_p$ suppress ordinary bulk semiclassical corrections. This is only a scale-separation statement, not a proof of semiclassical stability.

There is also a regularity caveat. The fiducial metric is $C^2$ but not $C^3$ at the bubble center, which is sufficient for a finite and continuous classical Riemann tensor but is not automatically sufficient for standard Hadamard point-splitting and renormalized-stress-tensor constructions, whose local counterterms involve higher derivatives of the geometry. An explicit RSET analysis should therefore either establish that the required renormalization framework remains well defined for the adopted profile or replace the central profile by a $C^\infty$ completion that preserves the same large-scale and wall properties.

This distinction is particularly important for superluminal warp geometries. Hiscock \cite{Hiscock1997} found severe quantum effects in an eternal two-dimensional Alcubierre background, while Finazzi, Liberati, and Barcel\'o \cite{Finazzi2009} studied dynamical warp drives and found a Hawking-like flux together with an exponentially growing renormalized stress-energy tensor near the front wall in their effective treatment. The modified propagator \eqref{eq:propagator} changes the ultraviolet structure entering such calculations, but no result obtained here demonstrates that it removes the relevant state-dependent or horizon-related growth. Establishing this would require an explicit computation of $\langle T_{\mu\nu}\rangle_{\rm ren}$ from a T-duality-consistent two-point function.

For $\el\sim\ell_p$, even the generic suppression factor in Eq.~\eqref{eq:backreaction} disappears. The minimal-profile-scale endpoint should therefore be regarded only as evidence that the classical geometry remains finite, not as a regime in which semiclassical dynamics is necessarily trustworthy.

\subsection{Quantum energy inequalities}

A complementary constraint arises from quantum energy inequalities, which bound suitable averages of negative renormalized energy densities. For a massless scalar field in four-dimensional Minkowski spacetime, Ford and Roman obtained the bound \cite{FordRoman1996,FordRoman1995}
\begin{equation}
\frac{\tau_0}{\pi}\int_{-\infty}^{\infty}
\frac{\langle T_{\mu\nu}u^\mu u^\nu\rangle}
{t^2+\tau_0^2}\,\dd t
\ge -\frac{3}{32\pi^2\tau_0^4},
\label{eq:FR}
\end{equation}
for the Lorentzian sampling function used in their analysis. More general quantum energy inequalities for different sampling functions, field theories, and curved backgrounds were developed subsequently; see, e.g., Refs.~\cite{FewsterEveson1998,FewsterSmith2008} and the review \cite{KontouSanders2020}. In particular, Ref.~\cite{FewsterSmith2008} establishes an absolute QEI for a minimally coupled massive scalar field on four-dimensional globally hyperbolic curved spacetimes. These results emphasize that the admissible magnitude and duration of negative energy depend on the quantum field theory, spacetime geometry, state, and sampling procedure, rather than on the classical stress tensor alone.

Pfenning and Ford \cite{PfenningFord1997} applied such reasoning to the Alcubierre geometry and found that, under their assumptions, a macroscopic bubble requires an extraordinarily thin wall. The present construction does not invalidate that conclusion. What changes is the ultraviolet input: the propagator \eqref{eq:propagator} is exponentially softened at momenta of order $1/\lo$, whereas standard QEI derivations rely on the usual short-distance/Hadamard structure of the field two-point function.

It would therefore not presently be justified either to apply the standard inequality unchanged at arbitrarily short sampling times or to replace $\tau_0$ by an ad hoc ``effective'' sampling scale without derivation. We make neither assumption here. The appropriate open problem is instead to construct the T-duality-modified two-point function on the warp background, determine its admissible state space and short-distance singularity structure, and derive the corresponding averaged-energy inequality. Only then can one decide whether the thick-walled macroscopic branch of Eq.~\eqref{eq:shape} can be supported by a consistent quantum state.

These two issues are closely linked. The exact energy \eqref{eq:totalenergy} characterizes the classical effective geometry, but neither its finiteness nor the suppression of bulk corrections in Eq.~\eqref{eq:backreaction} demonstrates that a quantum source realizing the required stress-energy tensor exists. Semiclassical backreaction and quantum energy inequalities therefore provide complementary consistency tests of the same effective construction.

\section{Discussion and conclusions}
\label{sec:discussion}

We have constructed and analyzed an Alcubierre-type warp geometry motivated by the zero-point length associated with string T-duality. The central ansatz identifies the warp profile with the complementary cumulative mass fraction of the T-duality-smeared source and introduces the effective one-scale interpolation $\el^2=R^2+\lo^2$. Once this prescription is adopted, the wall slope, Eulerian energy density, volume-integrated negative energy, York expansion, and axial causal diagnostics follow in closed form. In particular,
$E=-(15\pi/1024)\vs^2\el$, and the profile-width limit $R\to0$ remains finite at fixed $\vs$, with $\el\to\lo$ and $|E|\to(15\pi/1024)\vs^2\lo$. Thus the adopted minimal profile scale removes the divergences associated with shrinking the one-scale configuration toward its minimum profile scale $\lo$, without altering the requirement of exotic stress energy. This is distinct from the conventional fixed-radius thin-wall limit, which is not contained in the present one-scale ansatz.

The interpretation of this result is necessarily more limited than the algebraic regularity itself. The mapping from the cumulative T-duality mass distribution to a warp shape function is a phenomenological prescription, and Eq.~\eqref{eq:ell} is an effective interpolation rather than a relation derived uniquely from the T-duality propagator. In particular, the limit $\lo\to0$ recovers the smooth profile \eqref{eq:classicalprofile}, not an arbitrarily thin Alcubierre wall. The absence of a divergence in the one-scale contraction limit is therefore partly built into the ansatz and should not be read as a regularization of the independent fixed-radius thin-wall limit. A direct smearing of a finite-radius Alcubierre shell, designed to recover a prescribed classical profile as $\lo\to0$ while generating a calculable minimum wall thickness for $\lo>0$, would provide the natural next test of the underlying physical idea.

The generalized family \eqref{eq:family} reinforces this smooth one-scale character. At fixed $\el$ the transition region moves outward as $\sqrt n$ rather than approaching a top hat. The formal peak-fixed rescaling $\el_n\propto n^{-1/2}$ yields the smooth limiting profile \eqref{eq:familylimit} and the finite energy \eqref{eq:Efamilylimit}, but at fixed nonzero $\lo$ it cannot be continued to arbitrarily large $n$ because $\el_n\ge\lo$ imposes the bound \eqref{eq:nmax}. The family therefore does not conceal a distributional Alcubierre limit under the rescalings considered here.

Exotic stress energy remains unavoidable. Equation~\eqref{eq:density} gives $T_{\mu\nu}n^\mu n^\nu<0$ throughout the off-axis wall, so the weak energy condition is violated directly for the Eulerian observers. This is consistent with the general superluminal obstruction of Olum \cite{Olum1998} and with the Alcubierre/Nat\'ario analysis of Lobo and Visser \cite{LoboVisser2004}. Recent constructions have emphasized positive Eulerian energy densities or positive-energy source sectors \cite{BobrickMartire2021,Lentz2021,FellHeisenberg2021}; however, Santiago, Schuster, and Visser \cite{Santiago2022} stressed that positivity for a preferred observer family does not by itself establish the weak energy condition. The present construction therefore regularizes the magnitude and distribution of the required negative energy within the adopted family; it does not evade the standard energy-condition problem.

The causal analysis likewise requires a careful distinction between the full $3+1$ geometry and its axial reduction. For constant superluminal velocity, the exact solution \eqref{eq:rh_exact} of $\beta=1$ defines a spherical stationary-limit (Killing-norm-zero) surface of the comoving Killing field, not a spherical null horizon, as shown explicitly by Eq.~\eqref{eq:stationarynotnull}. On the symmetry axis it marks two stationary null-characteristic points with opposite linearized peeling behavior, in accord with the familiar front/rear causal asymmetry of superluminal Alcubierre geometries \cite{Clark1999}. The magnitude \eqref{eq:kappa_exact} is therefore interpreted only as an axial peeling scale. A genuine global horizon analysis requires solving the full null-hypersurface problem; until that is done, neither a $3+1$ Hawking temperature nor an area entropy should be assigned.

As emphasized in Sec.~\ref{sec:quantum}, regularity of the classical effective metric is not equivalent to semiclassical consistency. The scale estimate \eqref{eq:backreaction} suggests suppression of ordinary bulk quantum corrections for $\el\gg\ell_p$ at fixed velocity, but known renormalized-stress-tensor analyses \cite{Hiscock1997,Finazzi2009} show that special causal regions can dominate. Similarly, the ultraviolet softening of the propagator \eqref{eq:propagator} motivates a re-examination of quantum energy inequalities but does not determine their modified form. A T-duality-consistent two-point function, renormalized stress tensor, and averaged-energy bound are therefore needed before one can decide whether the thick-walled macroscopic branch can be supported by an admissible quantum state.

Superluminal geometries also retain the usual chronology concerns. Suitable combinations of warp-drive trajectories can generate closed causal curves \cite{Everett1996}, while Hawking's chronology-protection proposal \cite{Hawking1992} suggests that quantum backreaction may obstruct the formation of chronology horizons. Whether the presence of a zero-point length materially alters this conclusion remains an open question.

Several extensions follow naturally from the present analysis. Besides the direct finite-radius smearing problem, it would be useful to solve the full $3+1$ null-hypersurface equation and compute the renormalized stress tensor using the T-duality-modified propagator, thereby testing whether the instabilities found in Refs.~\cite{Hiscock1997,Finazzi2009} persist. Deriving, rather than postulating, the corresponding quantum energy inequality would then determine whether the exact thick-walled configurations obtained here can be supported by a consistent quantum state. It would also be instructive to repeat the construction in the zero-expansion Nat\'ario class \cite{Natario2002} and to compare systematically with modern positive-Eulerian-energy warp geometries \cite{Santiago2022,BobrickMartire2021,Lentz2021,FellHeisenberg2021} and with the recent general classification and no-go analysis of Ref.~\cite{Barzegar2026}.

Within its stated scope, the principal conclusion is therefore precise: a T-duality-motivated minimal profile scale can render an Alcubierre-type one-scale geometry finite and analytically tractable, but it does not remove exotic matter and does not by itself establish dynamical or quantum realizability. The regularization is explicit; the deeper causal, semiclassical, and quantum-consistency questions remain open.

\acknowledgments{
FSNL acknowledges support from the Funda\c{c}\~{a}o para a Ci\^encia e a Tecnologia (FCT) Scientific Employment Stimulus contract with reference CEECINST/00032/2018, and funding through the research grant UID/04434/2025.}

\end{document}